\documentclass[12pt,a4paper]{article}

\usepackage{authblk}

\usepackage{booktabs}
\usepackage{mathptmx}

\usepackage{amsmath,amsfonts,latexsym,amssymb}
\usepackage{verbatim,amsthm,curves,graphics}
\usepackage{mathrsfs}
\usepackage[dvips]{graphicx}
 
\usepackage{todonotes,soul} 
 
\numberwithin{equation}{section}
 
\usepackage{latexsym}
\usepackage{mathrsfs} 
\usepackage{epsfig}
\usepackage{longtable}
\usepackage{slashed}
\usepackage[all]{xy}
\usepackage{graphicx,tikz}
\usetikzlibrary{arrows}
\usepackage{setspace}

\usepackage{graphicx,color}
\usepackage{array}
\usepackage{tikz}
\usetikzlibrary{shapes.geometric}
\usetikzlibrary{decorations.pathmorphing}
\usetikzlibrary{arrows,shapes,positioning}
\usetikzlibrary{decorations.markings}
\tikzstyle arrowstyle=[scale=1]
\tikzstyle directed=[postaction={decorate,decoration={markings,mark=at position .65 with {\arrow[arrowstyle]{stealth}}}}]
\tikzstyle reverse directed=[postaction={decorate,decoration={markings,mark=at position .65 with {\arrowreversed[arrowstyle]{stealth};}}}]
\def\ben{\begin{equation}}
\def\een{\end{equation}}
\def\bena{\begin{eqnarray}}
\def\eena{\end{eqnarray}}

\def\d{{\rm d}}
\def\V{{\cal V}}

\def\F{{\cal F}}

\def\cP{{\mathcal P}}

\def\path{\operatorname{Pa}}

\def\V{\mathcal{V}}
\def\K{\mathcal{K}}
\def\F{\mathcal{F}}
\def\H{\mathcal{H}}

\newcommand{\gA}{{\mathfrak A}}

\newcommand{\RR}{{\mathbb R}}

\newcommand{\CC}{{\mathbb C}}

\renewcommand{\r}{\mathbf{\hat r}}

\newcommand{\dd}{{\operatorname{d}}}

\begin{document}
\title{News vs Information}
\author[1]{Stefan Hollands\thanks{\tt stefan.hollands@uni-leipzig.de}}
\affil[1]{Institute for Theoretical Physics, University of Leipzig,
	Br\"{u}derstra{\ss}e 16,
	D-04103 Leipzig, Germany.} 
\author[2]{Akihiro Ishibashi\thanks{\tt akihiro@phys.kindai.ac.jp}}
\affil[2]{Department of Physics, Kindai University, 
	Higashi-Osaka, 577-8502, Japan.} 
\date{27 March 2019}
\maketitle 

\begin{abstract}
We consider the relative entropy  between the vacuum state and a coherent state in linearized quantum gravity
around a stationary black hole spacetime. Combining recent results by Casini et al. and Longo with the Raychaudhuri 
equation, the following result is obtained: Let $\gA$ be the algebra of observables assoiciated with a region that is 
the causal future of some compact set in the interior of the spacetime. Let $S$ be the relative entropy with respect to this algebra, 
$A$ the area of the horizon cross section defined by the region, computed to second order in the gravitational perturbation. If 
the region is time-translated by the Killing parameter $t$, then $\frac{\dd}{\dd t}(S+A/4)= 2\pi F$, with $F$ the flux of the gravitational/matter radiation (integrated squared news tensor) 
emitted towards the future of the region.
\end{abstract}

\section{Introduction}

Information theoretic considerations in quantum field theory have attracted a lot of attention in recent years, not least due to intriguing relations 
with quantum field theory in curved spacetime or even (quantum) gravity theory, for instance through the ``quantum focussing conjecture'', its relation with Bekenstein bounds \cite{Longo1}, the ``quantum null energy condition'' \cite{bousso1,bousso2}, ``c-theorems'' \cite{Casini2} and many other topics. See e.g. \cite{Wall:2018ydq} for a survey with many references. See e.g. the book \cite{Hubeny} for an exposition of holographic ideas in this context.

In this short note we study the relative entropy between a vacuum like state and a coherent state (defined by a classical perturbation) in linearized quantum gravity around an asymptotically flat stationary vacuum black hole spacetime. The example we shall stick to for simplicity\footnote{All our arguments would go through also, e.g., for the Kerr spacetime, with no additional terms in the final formula related to rotation. 
The only difference besides a more complicated notation is that the results we rely on for the decay of perturbations are 
not available.} is the Schwarzschild metric (in $d>3$ dimensions)
given by 
\ben
\label{1}
\dd s^2=-[1- (r_0/r)^{d-3}]\dd t^2 + [1- (r_0/r)^{d-3}]^{-1} \dd r^2 + r^2 \dd \Omega (\r)^2_{S^{d-2}}, 
\een
where $\r \in S^{d-2}$.
We can consider, as usual, the ``tortoise'' coordinate defined by $\dd r_* = 
[1- (r_0/r)^{d-3}]^{-1} \dd r$. Then $v=\exp[\kappa(t+r_*)]$ is an affine parameter along the future horizon, $H^+$, and $u=t-r_*$ is a 
Bondi-type retarded time-coordinate at future null infinity, $I^+$. (Here $\kappa= (d-3)/({2}r_0)$ is the surface gravity.) 
Consider the future $H^+(v_0)$ of a cut at $v_0$ of the horizon consisting of the points on the 
horizon having $v$-coordinate greater than the value $v_0$. Similarly, consider the future $I^+(u_0)$ of a cut at $u_0$ of future null-infinity consisting of the points on the future null-infinity having $u$-coordinate greater than the value $u_0$. Let $\gA(u_0, v_0)$ be the algebra of observables for the linearized gravitational field 
associated with the ``domain of dependence'' of these regions, 
$D(u_0,v_0)=D^-( H^+(v_0) \cup i^+ \cup I^+(u_0))$ with $i^+$ future timelike infinity. See fig. \ref{Duv}. 
\usetikzlibrary{decorations.pathmorphing}
\tikzset{zigzag/.style={decorate, decoration=zigzag}}
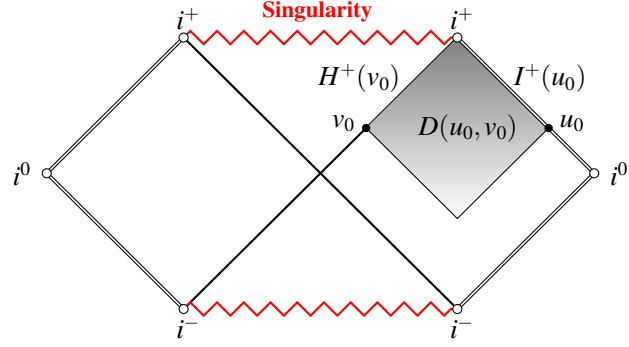
\begin{figure} 
\begin{center}
\begin{tikzpicture}[scale=0.6, transform shape]
\draw[thick](3,-3)--(0,0)--(3,3);
\draw[thick](-3,3)--(0,0)--(-3,-3);

\shade[top color=gray] (1,1) -- (3,3) -- (5,1)  -- (3,-1) -- cycle;
\node[anchor=west] at(2,1) {{\Large $D(u_0,v_0)$}};

\draw[double](3,3)--(6,0)--(3,-3);
\draw[double](-3,3)--(-6,0)--(-3,-3);
\draw[red,thick,zigzag] (-3,3) -- (3,3); 
\draw[red,thick,zigzag] (-3,-3) -- (3,-3); 
\draw(5,1)--(3,-1)--(1,1);

\node[anchor=south] at(0,3.2) {{\large {\color{red} {\bf Singularity} }}}; 

\draw (5,1) node[draw,shape=circle,scale=0.4,fill=black]{};
\node[anchor=west] at(4.1,2.1) {{\Large $I^+(u_0)$}};
\node[anchor=west] at(5.1,1.1) {{\Large $u_0$}};
\draw (1,1) node[draw,shape=circle,scale=0.4,fill=black]{};
\node[anchor=east] at(1.9,2.1) {{\Large $H^+(v_0)$}};
\node[anchor=east] at(0.9,1.1) {{\Large $v_0$}};

\node[anchor=north] at(3.1,3.8){{\Large $i^+$}};
\node[anchor=west] at(6.2,0) {{\Large $i^0$}};
\node[anchor=south] at(3.1,-3.8){{\Large $i^-$}};

\node[anchor=north] at(-2.9,3.8){{\Large $i^+$}};
\node[anchor=east] at(-6.2,0) {{\Large $i^0$}};
\node[anchor=south] at(-2.9,-3.8){{\Large $i^-$}};

\draw (3,3) node[draw,shape=circle,scale=0.5,fill=white]{};
\draw (6,0) node[draw,shape=circle,scale=0.5,fill=white]{};
\draw (3,-3) node[draw,shape=circle,scale=0.5,fill=white]{};

\draw (-3,3) node[draw,shape=circle,scale=0.5,fill=white]{};
\draw (-6,0) node[draw,shape=circle,scale=0.5,fill=white]{};
\draw (-3,-3) node[draw,shape=circle,scale=0.5,fill=white]{};
\end{tikzpicture}
\end{center}
\caption{
\label{Duv}
Conformal diagram of the maximally extended Schwarzschild spacetime and the past domain of dependence 
$D(u_0,v_0)$ of a partial Cauchy surface $H^+(v_0) \cup i^+ \cup I^+(u_0)$. 
}
\end{figure}
Furthermore, let $\omega_0$ be a state which is defined to be the vacuum with respect to the ``modes going through future null-infinity'' 
and a thermal state at the Hawking temperature $T=\kappa/2\pi$  
with respect to ``modes going through the future horizon''. Finally, consider a smooth, classical 
solution $h_{ab}$ to the linearized Einstein equations whose initial data are compactly supported within $D(u_0,v_0)$. 

Associated with such a solution, we can define a coherent state, $\omega_h$ of the linearized gravitational field. Then, we can consider the relative entropy $S(\omega_0/\omega_h)$ with respect to the 
partial observable algebra $\gA(u_0, v_0)$. It is the amount of information which an observer having access to all observables in $\gA(u_0, v_0)$ gains if she/he updates her/his belief about the system from state $\omega_0$ (black hole with no radiation) to $\omega_h$ (black hole with radiation).

The relative entropy between the vacuum state and a coherent state 
has also recently appeared in works by Casini et al. \cite{Casini} and Longo \cite{Longo} in the context of a free scalar field, and in \cite{hollands1} in the case of 1+1 conformal field theories. Here we apply their results (with some 
fairly trivial modifications) to the linearized gravitational field. Our main point is that the resulting formula can be rewritten in a suggestive form using 
Raychaudhuri's equation. 

To frame our result, we imagine that we have a 1-parameter family of classical solutions to the
 full vacuum Einstein equations, $g_{ab}(\lambda)$, such that $g_{ab}|_{\lambda=0}$ is 
the Schwarzschild background \eqref{1}, such that $\dd g_{ab}/\dd \lambda |_{\lambda=0}=h_{ab}$ is the given linear perturbation, such that the second order perturbed expansion $\dd^2 \theta/\dd \lambda^2|_{\lambda=0}$, goes to zero when $v \to \infty$.
By the Lemma 1 of \cite{hw}, we can -- and should -- assume that the linearized perturbation $h_{ab}$ is in a gauge such that 
the first order perturbed expansion, $\dd \theta/\dd \lambda |_{\lambda=0}=0$ on $H^+$ vanishes\footnote{This condition means that we are 
correctly identifying the coordinate location of the horizon to first order.}. 

Now we make a time translation by $t>0$, thereby moving the region $D(u_0,v_0)$ to $D(u_0+t,v_0 e^{\kappa t})$ towards the future. Then the relative entropy, 
which depends on the algebra $\gA(u_0,v_0)$ associated with this region, also changes -- in fact it has to decrease by the well-known monotonicity of this quantity. 
To emphasize this, we write the relative entropy also as $S(u_0,v_0)$ by abuse of notations.
Actually, we shall show that\footnote{We use units where $G=c=\hbar=1$.}
\ben
\label{claim}
\left( S(u_0,v_0) + \frac{\delta^2 A(v_0)}{4} \right)' = 2\pi \delta^2 \F(u_0) 
\een
where a prime $'=\frac{\d}{\d t}$. We also use the notation $\delta g_{ab}=\dd g_{ab}/\dd \lambda |_{\lambda=0}, \delta^2 g_{ab} = \frac{1}{2} \dd^2 g_{ab}/\dd \lambda^2|_{\lambda=0}$ etc., and\footnote{For coupled Einstein-scalar field theory, there would be a contribution to the flux $\F$ from the matter fields.}
\ben
\F(u_0)= -\frac{1}{32\pi} \int_{I^+(u_0)} N_{ab} N^{ab}\,  \dd u \dd^{d-2} \r
\een
where $N_{ab}$ is the news-tensor at $I^+$ (see \cite{hi,hiw,ht} for the precise definitions). The quantity $S+A/4$ on the left side is a version of Bekenstein's generalized entropy, and the derivative of it with respect to a time parameter also appears in the quantum focussing condition \cite{bousso1}. The quantity on the right side is the flux of gravitational radiation through $I^+(u_0)$.

\section{Relative entropy of coherent states of the Weyl algebra}
\label{sec:Weyl}
We first recall the definition of the Weyl algebra in an abstract setting adapted to the ``null quantization'' of the canonical commutation relations, for details see e.g. 
\cite{Kay&Wald, Moretti2, hwreview, Wald1}. This setting will be applied to both the future horizon and future null infinity below. To start, we define $\V_0=C_{0}^\infty(\RR \times S^{d-2}, \RR)$, to be viewed as the space of null initial data for a wave equation. On $\V$, a symplectic form is defined by $(\r \in S^{d-2})$
\ben
\label{symp1}
w(F_1,F_2)= \int_\RR \int_{S^{d-2}} (F_1 \partial_u F_2-F_2 \partial_u F_1)\, \dd u \dd^{d-2} \r . 
\een
The Weyl-algebra is defined by the relations $W(F_1)W(F_2) = \exp[-iw(F_1,F_2)/2] W(F_1+F_2)$ and $W(F)^*=W(-F)=W(F)^{-1}$. This algebra is next represented in a bosonic Fock space as follows. The 1-particle space is defined to be $\H_1= L^2(\RR_+ \times S^{d-2})$. An $F \in \V_0$ is mapped to $\H_1$
by projecting onto the ``positive frequency part'' $KF(k,\r):=\sqrt{2k} \hat F(k,\r)$, where a hat denotes the Fourier transform\footnote{Our convention for the Fourier transform is $\hat F(k)= \int e^{iku} F(u) \dd u$.} of $F(u,\r)$ 
in the variable $u$. 
On the bosonic Fock space $\H=\CC \oplus \oplus_{n=1}^\infty \H_n,$ with $\H_n$ the $n$-fold symmetrized tensor product of $\H_1$, 
we define bosonic creation operators by 
$a(k,\r)^*$ in the usual way (in the sense of distributions), with relation $[a(k_1,\r_1),a(k_2,\r_2)^*]=\delta(k_1-k_2)\delta^{d-2}(\r_1,\r_2)$, and then we represent the Weyl operators by 
the unitary operators
\ben
W(F)=\exp\left(a(KF)-a(KF)^* \right)
\een
on $\H$
with $a(\psi)=\int_0^\infty \int_{S^{d-2}}a(k) \psi(k) \dd k \dd^{d-2} \r $. For the vacuum state $|\Omega_0\rangle$ in Fock space, one then gets the formula $\langle \Omega_0 | W(F) \Omega_0\rangle = \exp (-\|KF\|^2/2)$, where we mean the $L^{2}$-norm on the right side. In fact, we can write this $L^2$-inner product also as 
\ben
\label{Inner}
(KF_1,KF_2) = \frac{-1}{\pi} \int \frac{F_1(u_1,\r) F_2(u_2,\r)}{(u_1-u_2-i0)^2} \dd u_1 \dd u_2  \dd^{d-2} \r .
\een
Since $F_1,F_2$ are real-valued,
we have $\Im (KF_1,KF_2) = w(F_1,F_2)/2$, which can be used to show that the formulas are consistent with the Weyl relations. 
For any $u_0$, we define the algebra $\gA(u_0)$ to be the weak closure\footnote{In other words, the double commutant.} 
of $\{W(F) \mid F \in \V_0, {\rm supp} F \subset (u_0,\infty) \times S^{d-2} \}$. 

We next recall the definition of the relative entropy in terms of modular operators due to Araki. For details on 
operator algebras in general and references we refer to \cite{Bratteli} and for a recent survey of operator algebraic methods in quantum information theory in
QFT, we refer to \cite{sanders_2}. A nice exposition directed towards theoretical physics audience is \cite{Witten:2018zxz}.

Let $\gA$ be a v. Neumann algebra\footnote{An algebra of bounded operators that is 
closed in the topology induced by the size of matrix elements. The review of the expository material until eq. \eqref{Srel} follows \cite{hollands1}.} of 
operators on a Hilbert space\footnote{We always assume that $\H$ is separable.}, $\H$. We assume that $\H$ contains a ``cyclic and separating'' vector for $\gA$, that is, a unit vector $|\Omega \rangle$
such that the set consisting of $X|\Omega\rangle$, $X \in \gA$ is a dense subspace of $\H$, and such that $X|\Omega\rangle=0$ always implies $X=0$
for any $X \in \gA$. We say in this case that $\gA$ is in ``standard form'' with respect to the given vector. $\gA^+$ denotes the set of positive, self-adjoint elements in $\gA$ (which are always of the form $X=Y^*Y$ for some $Y \in \gA$).

In this situation, one can define the Tomita operator $S$ on the domain ${\rm dom}(S) = \{ X|\Omega\rangle \mid X \in \gA\}$ by 
\ben
SX|\Omega \rangle = X^* |\Omega \rangle
\een
The definition is consistent due to the cyclic and separating property. It is known that $S$ is a closable operator, and we denote its closure by the same symbol. This closure has a polar decomposition denoted by $S=J\Delta^\frac12$, with $J$ anti-linear and unitary and $\Delta$ self-adjoint and non-negative. 
Tomita-Takesaki theory concerns the properties of the operators $\Delta, J$. The basic results of the theory are the following, see e.g. \cite{Bratteli}: 
\begin{enumerate}
\item 
$J \gA J = \gA'$, where the prime denotes the {\bf commutant} (the set of all bounded operators on $\H$ 
commuting with all operators in $\gA$) and $J^2 = 1, J\Delta J = \Delta^{-1}$, 

\item 
If $\alpha_t(X) =  \Delta^{it} X \Delta^{-it}$, then $\alpha_t \gA = \gA$ and $\alpha_t \gA' = \gA'$ for all $t \in \RR$, $\Delta^{it}|\Omega\rangle=|\Omega\rangle$
for all $t \in \RR$.

\item
The positive, normalized (meaning $\omega(X) \ge 0 \, \, \forall X \in \gA^+, \omega(1) = 1$) 
linear {\bf expectation functional} 
\ben 
\omega(X) = \langle \Omega |X\Omega \rangle
\een
satisfies the {\bf KMS-condition} 
relative to $\alpha_t$. This condition states that for all $X,Y \in \gA$, the bounded function 
\ben
\label{Fdef}
t \mapsto F_{X,Y}(t) = \omega (X \alpha_t(Y)) \equiv \langle \Omega | X \Delta^{it} Y \Omega \rangle
\een
has an analytic continuation to the strip $\{ z \in \CC \mid -1 < \Im z < 0 \}$ with the property that its boundary value for $\Im z \to -1^+$ exists and
is equal to
\ben
\label{KMS}
F_{X,Y}(t-i) = \omega (\alpha_t(Y) X).
\een

\item 
Any normal (i.e. continuous in the weak$^*$-topology) positive linear functional $\omega'$ on $\gA$ has a unique vector representative 
$|\Omega' \rangle$ in the natural cone 
\ben
\cP^\sharp = \overline{\{ \Delta^{1/4} X |\Omega \rangle \mid 
X \in \gA^+\}}=
\overline{\{ Xj(X) |\Omega \rangle \mid X \in \gA\}}, 
\een
where the overbar means closure and $j(X) = JXJ$. The state functional is thus
$\omega'(X) = \langle \Omega' |X\Omega' \rangle$ for all $X \in \gA$. 
\end{enumerate}

\medskip
\noindent

A generalization of this construction is that of the relative modular operator, flow etc. For this purpose, let $\omega'$ be a normal state on $\gA$, $|\Omega' \rangle$ its unique vector representative in the natural cone in $\H$, which is assumed (for simplicity) to be cyclic and separating, too. 
Then we can consistently define
\ben
S_{\omega, \omega'} X |\Omega' \rangle = X^* |\Omega \rangle
\een
form the closure, and make the polar decomposition $S_{\omega, \omega'}=J_\omega^{} \Delta^\frac12_{\omega,\omega'}$. 
The {\bf Araki relative entropy} is defined by 
\ben
S(\omega / \omega') = \langle \Omega | (\log  \Delta_{\omega,\omega'}) \Omega \rangle. 
\een
In the case of Type I factors (``quantum mechanics''), e.g. $\gA=M_N(\CC)$, the situation is this: State functionals are equivalent to 
density matrices $\rho$ via $\omega(X) = {\rm Tr}(X\rho)$, $\H$ is  the algebra itself $M_N(\CC) \cong \CC^N \otimes \CC^N$ on which $\gA$ acts by left multiplication. 
The state $|\Omega\rangle$ corresponds to $|\rho^{1/2}\rangle$, the inner product is $\langle X | Y \rangle = {\rm Tr}(X^* Y)$, the modular operator is
$\Delta = \rho \otimes \rho^{-1}$, and the relative modular operator is $\rho \otimes \rho^{\prime -1}$ , where $\rho'$ is the density matrix associated with $\omega'$. Using this, one verifies
$S(\omega / \omega') = {\rm Tr} \rho ( \log \rho - \log \rho')$. These formulae do not hold for 
type III factors which occur in quantum field theories.

The relative entropy has many beautiful properties.  It is e.g. never negative, but can be infinite, is decreasing under completely positive maps, is jointly convex in both arguments, etc. The physical interpretation of $S(\omega / \omega')$ is the amount of information gained if we update our belief about the system from the state $\omega$ to $\omega'$.

In this paper, we are interested in the special case when $\omega'=\omega_U$, where
\ben
\omega_U(X) \equiv \omega(U^* X U) = \langle U \Omega | XU\Omega \rangle,  
\een
and where $U$ is some unitary operator from $\gA$. 
The corresponding vector representative in the natural cone is $|\Omega_U \rangle=Uj_\omega (U)|\Omega \rangle$, with $j_\omega (X)= J_\omega X J_\omega$. 
Going through the definitions, one finds immediately that $j_\omega(U) \Delta^{1/2}_\omega j_\omega(U^*) = \Delta^{1/2}_{\omega, \omega'}$, implying that
\ben\label{Srel}
S(\omega/  \omega_U) = -\langle U^* \Omega |  (\log \Delta) U^*\Omega \rangle = i\frac{\dd}{\dd t} \langle U^* \Omega |  \Delta^{it} U^*\Omega \rangle \Bigg|_{t=0}, 
\een
where $\Delta$ is the modular operator of the original state $\omega$.

Now we specialize to the following situation: $\gA=\gA(u_0)$ is the v. Neumann closure of the partial Weyl algebra
$\{W(F) \mid {\rm supp} F \subset (u_0,\infty) \times S^{d-2} \}$. $|\Omega\rangle=|\Omega_0\rangle$ is the vacuum. This state is cyclic and separating 
by the Reeh-Schlieder theorem, and therefore the modular operator exists. $U=W(F)^*=W(-F)$ is a specific Weyl-operator for some $F \in \V_0$ supported in the interval 
$(u_0, \infty)$. We call the corresponding state functionals $\omega_0$ (vacuum) and $\omega_F(X)=\omega(W(F) X W(F)^*)$ (coherent state). We wish to compute 
$S(\omega_0 / \omega_F)$ relative to the algebra $\gA(u_0)$.

As one may expect, the general relation \eqref{Srel} can be expressed in the case at hand in terms of the 1-particle quantities only. This is done as follows.
Let $\K \subset \H_1$ be the image under $K$ of all possible $G \in \V_0$ such that the support of $G$ is contained in $(u_0, \infty)$. Since $\V_0$ is a real vector space, $\K$ is only a real linear subspace of $\H_1$. However, the 1-particle version of the Reeh-Schlieder theorem states that 
$\K + i\K$ is dense in $\H_1$ and $\K \cap i\K=0$. Such a real subspace is called ``standard''. For a standard real subspace of a Hilbert space one 
can define 1-particle versions of the Tomita-operators on the dense domain $\K + i\K$ by \cite{BrunettiGuidoLongo}
\ben
S_\K(\psi + i\phi) = \psi-i\phi, \quad \phi, \psi \in \K, \quad 
S_\K = J_{\K} \Delta_{\K}^{1/2}, 
\een
and these objects have properties analogous to those of the Tomita-operators for a v. Neumann algebra in standard form. In the case at hand, 
the 1-particle modular operator acts by dilations of the coordinate $u$ 
around the point $u_0$, more precisely 
\ben
\label{deltac}
\Delta^{it}_\K KG = K(G \circ \Lambda_t), \quad \Lambda_t(u,\r)=(u_0+e^{-2\pi t}(u-u_0), \r),  \qquad \forall G \in \V_0,
\een
by 
the 1-particle version of the Bisognano-Wichmann theorem \cite{bisognano}. Furthermore, the full modular operator $\Delta^{it}=\Gamma(\Delta_\K^{it})$
is the second-quantized version of the 1-particle modular operator. In fact, one can show in general \cite{Longo} (Proposition 3.1)
\ben
\label{Srel2}
\frac{\dd}{\dd t} \langle \Omega_0 | W(F)^* \Delta^{it} W(F) \Omega_0 \rangle \Bigg|_{t=0} = \frac{\dd}{\dd t} (KF, \Delta^{it}_\K KF) \Bigg|_{t=0},
\een
which when evaluating the right side gives in combination with eqs. \eqref{Inner}, \eqref{deltac} and \eqref{Srel} the relation
\ben
\label{Srel3}
S(\omega_0 / \omega_F) = 2\pi \int_{u_0}^\infty \int_{S^{d-2}} (u-u_0) \partial_u F(u,\r)^2 \, \dd u  \dd^{d-2} \r . 
\een

\section{Relative entropy for scalar and gravitational perturbations}

We now apply the abstract setting of the previous section to scalar (first) and then gravitational perturbations of Schwarzschild. Consider 
first a classical solution $\nabla^a \nabla_a \phi = 0$ to the massless Klein-Gordon (KG) equation with smooth, compactly supported initial data on some Cauchy surface $\Sigma$, i.e. 
a surface stretching from the bifurcation surface to spatial infinity such as the $t=0$ surface in the coordinates $(r_*,t,\r)$. 
The symplectic form $w$ is given by 
\ben
\label{symp0}
w(\phi_1, \phi_2) = \int_\Sigma (\phi_1 \nabla_a \phi_2 - \phi_2 \nabla_a \phi_1) \dd \Sigma^a
\een
and is independent of the choice of Cauchy surface by the usual argument based on Gauss' theorem. Now consider deforming 
$\Sigma$ to the degenerate ``surface'' consisting of the union of $H^+$, $I^+$ and (formally) future timelike infinity $i^+$, see fig. \ref{Duv}. 
By taking a suitable limit, one may 
hope that
\ben
\label{symp2}
w(\phi_1, \phi_2) = \int_{H^+} (F_1 \partial_v F_2-F_2 \partial_v F_1)\, \dd v \dd^{d-2} \r 
+ \int_{I^+} (F_1 \partial_u F_2-F_2 \partial_u F_1)\, \dd u \dd^{d-2} \r . 
\een
Here $\dd^{d-2} \r$ is the integration element of the unit radius round sphere in the case of $I^+$ and 
$r_0^{d-2}$ times that in the case of $H^+$.
$F$ denotes the restriction of $\phi$ to the future horizon in the case of $H^+$, and the restriction of $r^{d/2-1} \phi$ to future null infinity 
in the case of $I^+$. The latter restriction is immediately seen to exist by a standard conformal transformation of the Schwarzschild metric 
to an unphysical metric smooth at $I^+$. However, it is not obvious that the above integrals are absolutely convergent for $u,v \to \infty$ and that 
the expression \eqref{symp2} really agrees 
with the original definition \eqref{symp0} of the symplectic form over $\Sigma$, i.e. that there is no ``leakage of symplectic flux'' through $i^+$. This is shown in Thm. 2.2 of \cite{Moretti1} [which is using non-trivial pointwise decay results on $\phi$ due to \cite{Dafermos1,Luk} in $d=4$, and is extendable to $d>4$ in view of \cite{Schlue}]. More precisely, these authors establish that if one enlarges $\V_0$ to suitable subspaces $\V_H$ respectively $\V_I$ 
in each case so that it contains the restrictions of $\phi$ to $I^+$ respectively $H^+$, 
then the map $\phi \mapsto (\phi |_{H^+}, r^{d/2-1}\phi |_{I^+}) \in \V_H \times \V_I$, taking values in this enlarged space, is a symplectic and injective map.  In view of the mentioned decay results, it is possible e.g. to choose $\V_H$ for some $v_0>0$ to consist of those $F \in C^\infty(\RR \times S^{d-2};\RR)$ such that ${\rm supp}(F) \subset (v_0,\infty] \times S^{d-2}$ with a decay of the type $|F(v)| \le C(\log v)^{-3/2+\epsilon}$ as $v \to \infty$, 
for arbitrarily small $\epsilon>0$\footnote{$C$ depends on this $\epsilon$.} and with a decay of $|\partial_v F(v)|$ faster by a factor of $v^{-1}$. It is possible to choose $\V_I$ for some $u_0$ to consist of those $F \in C^\infty(\RR \times S^{d-2};\RR)$
such that ${\rm supp}(F) \subset (u_0,\infty]\times S^{d-2}$ for some $v_0>0$ with a decay of the type $|F(u)| \le Cu^{-1/2}$ for $u \to \infty$, with a decay of $|\partial_u F(u)|$ faster by a factor of $u^{-1/2}$.  

The second formula \eqref{symp2} for the symplectic form shows by comparison with \eqref{symp1} that one may quantize the theory by two copies of the Weyl algebra (for $I^+$ and $H^+$) as described in the previous section, now based on the symplectic space $\V_H \times \V_I$. Furthermore, one may define a ``vacuum'' state by making the above Fock space construction for both copies. See \cite{Moretti1} for a more detaild description of this procedure and especially Prop. 3.3(b) of \cite{Moretti1} for the precise definition of the map $K$ in this case.
Thereby, we get algebras of the form $\gA(u_0,v_0) := \gA_{H^+}(v_0) \otimes \gA_{I^+}(u_0)$, with $v_0>0, u_0 \in \RR$. Such an algebra is associated with the region $D(u_0,v_0)=D^-( H^+(v_0) \cup i^+ \cup I^+(u_0))$ as described in the introduction, see fig. \ref{Duv}. This von Neumann algebra is by construction large enough to contain the Weyl unitaries $W(F) \equiv W(F_{H^+}) \otimes W(F_{I^+})$ associated with the null data 
coming from the restriction map $\phi \mapsto (\phi |_{H^+}, r^{d/2-1}\phi |_{I^+})=F \in \V_H \times \V_I$ of a smooth solution $\phi$ with compact support on $\Sigma$, provided the causal future of the support of the initial data is within $D(u_0,v_0)$. 

Note that, owing to the relationship between Killing time $t$ and affine time $v$ on $H^+$, a time translation by $t$ corresponds to changing 
$v$ to $e^{\kappa t}v$. By the usual arguments (see e.g. \cite{Kay&Wald,hwreview}), the ``vacuum'' state defined on $\gA_{H^+}$ corresponds to a thermal state 
with respect to Killing time translations at the Hawking temperature $T=\kappa/2\pi$.\footnote{By the arguments of \cite{Thesis,Moretti1,Moretti2},
the state $\omega_0$ is a Hadamard state in the exterior region of the Schwarzschild spacetime, which in general becomes singular on $H^+$, by analogy with the Unruh-vacuum.} Furthermore, under a translation by Killing time $t$, 
our region $D(u_0,v_0)$ is moved into $D(u_0+t,e^{\kappa t} v_0)$. 

A classical solution to $\nabla^a \nabla_a \phi=0$ such that the causal future of the support of its initial data is contained in $D(u_0,v_0)$ 
gives rise to a coherent state on $\gA(u_0,v_0)$ by the construction of the previous section, taking $F=(\phi |_{H^+}, r^{d/2-1}\phi |_{I^+}) \in \V_H \times \V_I$  to be the restriction of $\phi$ to $I^+$ respectively $H^+$ (characteristic data), by taking $U=W(F)^*=W(F_{H^+})^* \otimes W(F_{I^+})^*=U_H \otimes U_I$. We call the coherent state $\omega_\phi$ in the present situation to emphasize its origin from the classical solution, $\phi$. 
To justify the name ``coherent state'' for that state, we recall (see e.g. \cite{Wald2}) that the quantum KG field $\Phi$ is related to the Weyl operators 
formally by $W(F)=\exp(iw(F,\Phi))$, where $F\in \V_H \times \V_I$ corresponds to the characteristic initial data on $H^+,I^+$ of some classical solution, and $w$ is as in 
eq. \eqref{symp2}. By the Weyl relations, one can then show, with $F$ the characteristic initial data of the classical solution $\phi$:
\ben
W(F) \Phi(x) W(F)^* = \Phi(x)+\phi(x)1,  
\een
so in particular $\omega_F(\Phi(x))=\omega_0(W(F)\Phi(x)W(F)^*)=\phi(x)$, as expected from a coherent state. 

By eqs. \eqref{Srel}, \eqref{Srel2}, \eqref{Srel3} applied to the two copies $\gA(u_0,v_0) := \gA_{H^+}(v_0) \otimes \gA_{I^+}(u_0)$ we get using the additivity of the relative entropy under the tensor product:
\ben
\label{Srelphi}
S(\omega_0 / \omega_\phi) = 2\pi \int_{I^+}  (u-u_0) \partial_u \tilde \phi(u,\r)^2 \, \dd u  \dd^{d-2} \r 
+ 2\pi \int_{H^+}  (v-v_0) \partial_v \phi(v,\r)^2 \, \dd v  \dd^{d-2} \r, 
\een
where $\tilde \phi = \lim_{I^+} r^{d/2-1}\phi$. At this stage, $S(\omega_0 / \omega_\phi)$ could still be infinite, but we now argue that the right side is actually finite. Finiteness of the second integral immediately follows from the 
characterization of the space $\V_H$ in which $F_{H^+}=\phi |_{H^+}$ lives. Finiteness of the first integral does not follow from the characterization of 
the space $\V_I$ where $F_{I^+}=r^{d/2-1}\phi |_{I^+}$ lives, but we can argue as follows, restricting for simplicity attention to $d=4$. Let $\F(u_1,u_2)=
\int_{u_1}^{u_2} (\partial_u \tilde \phi)^2 \dd u \dd^{2} \r$ be the flux through $I^+$ between $u_1,u_2$. By a result of \cite{Dafermos1}, Thm. 7.1 (18), 
$\F(u_1,u_2) \le C{\rm max} \{u_1, 1\}^{-2}$ for all $u_2 \ge u_1$. Let $\alpha>1$. The first integral in \eqref{Srelphi} is bounded above by (without loss of generality $u_0 \ge 1$) $\le \sum_{N=0}^\infty (N+1)^\alpha \F(u_0+N^\alpha,u_0+(N+1)^\alpha) \le C {\rm max} \{u_0^{-2}, 4^\alpha\} \sum_{N=0}^\infty (N+1)^{-\alpha}<\infty$.
 
A completely analogous analysis can be made in principle for the case of a gravitational perturbation $h_{ab}$, i.e. a smooth solution of the same type to the linearized vacuum Einstein equations. However, to make our analysis completely rigorous, we would need to justify the decay of gravitational perturbations at $I^+$ and 
$H^+$, for instance when proving the analog of \eqref{symp2}. Here one can use recent results by \cite{Dafermos2}. Since these results concern the Teukolski equation rather than linearized Einstein equation, one would additionally have to represent a gravitational perturbation in terms of Hertz-type potential solutions to Teukolsky's equation \cite{Chrzanowski,Kegeles}. Such an analysis would go beyond the scope of this short note, and one could, at any
rate, always consider solutions which, in some gauge, are smooth and of compact support at $I^+$ and $H^+$, obtained by a characteristic initial value problem
as considered e.g. in \cite{Friederich}.\footnote{Such solutions would not arise from 
compactly supported initial data in the interior, though.} We shall therefore proceed more formally, assuming that the integrals in questions converge, as they did in 
the case of the scalar field. 

The linearized Einstein equations are
$\nabla^c \nabla_c h_{ab} + \nabla_a \nabla_b h^c{}_c-\nabla_c\nabla_ah^c{}_b -\nabla_c\nabla_bh^c{}_a=0$. 
The symplectic forms on the pair of null surfaces $I^+$ respectively $H^+$ are given 
in eqs. (97) respectively (102) of \cite{hw}, which uses results of \cite{hi}, where we note that the boundary terms are absent in our setting.  Going through the analogous steps as for a scalar field and using formulas such as (96) and (101) of \cite{hw}, one finds 
\ben
\label{Srelh}
\begin{split}
\frac{1}{2\pi} S(\omega_0 / \omega_h) =& \frac{1}{32\pi} \int_{I^+(u_0)}  (u-u_0) \delta N_{ab} \delta N^{ab} \, \dd u  \dd^{d-2} \r \\
+ &\frac{1}{8\pi} \int_{H^+(v_0)}  (v-v_0) \delta \sigma_{ab} \delta \sigma^{ab} \, \dd v  \dd^{d-2} \r, 
\end{split}
\een
where $\delta N_{ab}$ is the perturbed news tensor of the perturbation $h_{ab}=\delta g_{ab}$ on $I^+$ \cite{hi}, and where $\delta \sigma_{ab}$ is the perturbed 
shear on $H^+$. 

On $H^+$, we now consider the Raychaudhuri equation (see e.g. \cite{Wald2}) for the 1-parameter family $g_{ab}(\lambda)$ described in the introduction, 
\ben
\frac{\dd}{\dd v} \theta(\lambda) = -\frac{1}{d-2} \theta(\lambda)^2 - \sigma_{ab}(\lambda) \sigma^{ab}(\lambda)- 8\pi \, T_{vv}(\lambda). 
\een
Taking two derivatives with respect to the family parameter $\lambda$, using that $\theta=\delta \theta=0$ on $H^+$ as well as 
$\sigma_{ab}=0$ on $H^+$ in the background, and using that the stress tensor vanishes in the present situation, 
we see $\frac{\dd}{\dd v}\delta^2 \theta = -\delta \sigma_{ab} \delta \sigma^{ab}$ on $H^+$. 
This formula is now used in the second term in eq. \eqref{Srelh}, giving 
\ben
\label{Srelh1}
\begin{split}
\frac{1}{2\pi} S(\omega_0 / \omega_h) =& \frac{1}{32\pi} \int_{I^+(u_0)}  (u-u_0) \delta N_{ab} \delta N^{ab} \, \dd u  \dd^{d-2} \r \\
 - & \frac{1}{8\pi} \int_{H^+(v_0)}  (v-v_0) \frac{\dd}{\dd v} \delta^2 \theta \, \dd v  \dd^{d-2} \r. 
\end{split}
\een
Finally, we time translate the region $D(u_0,v_0)$ by $t$ to $D(u_0+t,e^{\kappa t} v_0)$, take a derivative of \eqref{Srelh1} 
with respect to $t$, and use the relation $-\frac{\dd}{\dd v_0} \delta^2 A(v_0) = \int_{H^+(v_0)}  \frac{\dd}{\dd v} \delta^2 \theta \, \dd v  \dd^{d-2} \r$, which uses our assumption that 
$\delta^2 \theta(v) \to 0$ as $v \to \infty$. This then immediately gives the claim \eqref{claim} made in the introduction\footnote{ 
As mentioned before, the arguments given here would be straightforwardly generalized to Kerr spacetimes with no serious obstraction. 
For extremal Kerr spacetimes, the surface gravity vanishes, $\kappa=0$, and the horizon affine parameter is given by $v_0= t+ r_*$ with $r_*$ appropriately defined. 
In this case the time translation by $t$ should be taken simply as $D(u_0,v_0) \rightarrow D(u_0+t, v_0+t)$, and one would obtain the same claim \eqref{claim}. The formula \eqref{Srelh1} should also hold for stationary, asymptotically AdS black holes with the elimination of the first integral of right-hand side as the news $\delta N_{ab}$ vanishes due to the reflecting nature of 
the AdS boundary conditions. Then the formula \eqref{claim} represents simply a conservation of the generalized entropy, see \cite{Raamsdonk}
for similar looking formulas. }. 

We may use the same argument if we have a linear gravitational field and a massless scalar Klein-Gordon field. For classical solutions $h_{ab},\phi$ 
of the type described for both theories, we now get a coherent state $\omega_{h,\phi}$. The relative entropy with the reference state $\omega_0$
is given by the sum of \eqref{Srelh} and \eqref{Srelphi}. The second order Raychaudhuri equation now gives $\frac{\dd}{\dd v}\delta^2 \theta = -\delta \sigma_{ab} \delta \sigma^{ab} - 8\pi (\partial_v \phi)^2$ on $H^+$, and the flux $\F$ now also contains a contribution from the scalar field. With this contribution included, we 
get the same equation as stated in the introduction \eqref{claim}. 

Note that since the horizon area does not change to first and zeroth order in the classical perturbation $(h_{ab},\phi)$ defining the coherent state, and since 
the background news tensor is zero,
we may also write \eqref{claim} in the more suggestive form
\ben
(S+A/4)'=2\pi \F, 
\een
which holds up to and including second order in the perturbation, where $S$ is the relative entropy between the vacuum state and the coherent state. This equation relates an information theoretic quantity on the left side to the Bondi news tensor giving the flux on the right side.
It is tempting to speculate that this formula continues to hold in perturbative quantum gravity to all orders, and perhaps even for full quantum gravity. It would also be interesting to relate our formula to, e.g. \cite{8,9,10}.\footnote{Some other interesting connections between the news and information theoretic quantities may also be worth investigating. For instance, in \cite{hiw2}, we found a relation between the supertranslation, $T$ associated with a burst or radiation emitted by the collision of particles. Interestingly, $T$ can be written as 
\ben
T(\mathbf{\hat r}) = 2E\sum_{(i)\textnormal{ in, out}}\eta_{(i)} \rho_{(i)} \log \rho_{(i)}, \quad 
\rho_{(i)} = 
(E_{(i)}-\mathbf{\hat{r}\cdot p}_{(i)})/E
\een
where $E$ is the total energy of all particles, $\eta_{(i)}=\pm$ for in/outgoing particles, 
and $(E_{(i)},\mathbf{p}_{(i)})$ the four-momentum of particle $i$. Note that $\{\rho_{(i)}\}$ for the in- and outgoing momenta 
are two probability distributions (this follows from four-momentum conservation and the fact that $E_{(i)} \ge |\mathbf{p}_{(i)}|$). 
Furthermore, the right side has the form of $2E$ times the v. Neumann entropy difference 
of the in- and out probability distributions $\{\rho_{(i)}\}$.
Perhaps this is more than an accident.}

\medskip
\noindent
{\bf Acknowledgements:} Part of this work was carried out while S.H. was visiting IHES, Paris. It is a pleasure to thank IHES for hospitality and financial assistance. 
S.H. is grateful to the Max-Planck Society for supporting the collaboration between MPI-MiS and Leipzig U., grant Proj. Bez. M.FE.A.MATN0003. 
The work of A.I. was supported in part by JSPS KAKENHI Grants No.~15K05092.  



\begin{thebibliography}{99}

\bibitem{Longo1} 
  R.~Longo and F.~Xu,
  ``Comment on the Bekenstein bound,''
  J.\ Geom.\ Phys.\  {\bf 130}, 113 (2018)

\bibitem{bousso1} 
  R.~Bousso, Z.~Fisher, S.~Leichenauer and A.~C.~Wall,
  ``Quantum focusing conjecture,''
  Phys.\ Rev.\ D {\bf 93}, no. 6, 064044 (2016)


\bibitem{bousso2} 
  R.~Bousso, Z.~Fisher, J.~Koeller, S.~Leichenauer and A.~C.~Wall,
  ``Proof of the Quantum Null Energy Condition,''
  Phys.\ Rev.\ D {\bf 93}, no. 2, 024017 (2016)


    \bibitem{Casini2} 
  H.~Casini and M.~Huerta,
 `` A c-theorem for the entanglement entropy,''
  J.\ Phys.\ A {\bf 40}, 7031 (2007)

\bibitem{Wall:2018ydq} 
  A.~C.~Wall,
  ``A Survey of Black Hole Thermodynamics,''
  arXiv:1804.10610 [gr-qc].
  
  
  \bibitem{Hubeny}
M. Rangamani and T. Takayanagi, {\it Holographic Entanglement Entropy}, Springer Lecture Notes in Physics (2017)
  



\bibitem{Casini}
  H.~Casini, S.~Grillo and D.~Pontello,
  ``Relative entropy for coherent states from Araki formula,''
  arXiv:1903.00109 [hep-th].


\bibitem{Longo}  
  R.~Longo,
 ``Entropy of Coherent Excitations,''
  arXiv:1901.02366 [math-ph].
  
  \bibitem{hollands1}
 S.~Hollands,
  ``Relative entropy for coherent states in chiral CFT,''
  arXiv:1903.07508 [hep-th]. 
  
  
  \bibitem{hw}
   S.~Hollands and R.~M.~Wald,
  ``Stability of Black Holes and Black Branes,''
  Commun.\ Math.\ Phys.\  {\bf 321}, 629 (2013)
  
  \bibitem{hi}
  S.~Hollands and A.~Ishibashi,
  ``Asymptotic flatness and Bondi energy in higher dimensional gravity,''
  J.\ Math.\ Phys.\  {\bf 46}, 022503 (2005)
  
  \bibitem{hiw}
  S.~Hollands, A.~Ishibashi and R.~M.~Wald,
  ``BMS Supertranslations and Memory in Four and Higher Dimensions,''
  Class.\ Quant.\ Grav.\  {\bf 34}, no. 15, 155005 (2017)
  
  \bibitem{ht}
 S.~Hollands and A.~Thorne,
  ``Bondi mass cannot become negative in higher dimensions,''
  Commun.\ Math.\ Phys.\  {\bf 333}, no. 2, 1037 (2015)

\bibitem{Kay&Wald}
B.~S.~Kay and R.~M.~Wald,
  ``Theorems on the Uniqueness and Thermal Properties of Stationary, Nonsingular, Quasifree States on Space-Times with a Bifurcate Killing Horizon,''
  Phys.\ Rept.\  {\bf 207}, 49 (1991).

\bibitem{Moretti2}
C.~Dappiaggi, V.~Moretti and N.~Pinamonti,
  ``Hadamard States From Light-like Hypersurfaces,''
  SpringerBriefs Math.\ Phys.\  {\bf 25} (2017)

\bibitem{hwreview}
S.~Hollands and R.~M.~Wald,
  ``Quantum fields in curved spacetime,''
  Phys.\ Rept.\  {\bf 574}, 1 (2015)
  
  

\bibitem{Wald1}
R. M. Wald, {\it Quantum field theory in curved spacetime and black hole thermodynamics,}
U. of Chicago Press (1994)

  \bibitem{Bratteli}
 O.~Bratteli and D.~W. Robinson,
\newblock {\em {Operator Algebras and Quantum Statistical Mechanics I}}.
\newblock Springer (1987)
O.~Bratteli and D.~W. Robinson.
\newblock {\em {Operator Algebras and Quantum Statistical Mechanics II}}.
\newblock Springer (1997)
  
  \bibitem{sanders_2}
  S.~Hollands and K.~Sanders,
  \emph{Entanglement measures and their properties in quantum field theory,}
  SpringerBriefs in Mathematical Physics (2018)
  arXiv:1702.04924 [quant-ph].
  
  \bibitem{Witten:2018zxz} 
  E.~Witten,
  ``Notes on Some Entanglement Properties of Quantum Field Theory,''
  arXiv:1803.04993 [hep-th].
 
\bibitem{BrunettiGuidoLongo}
R.~Brunetti, D.~Guido and R.~Longo,
  ``Modular localization and Wigner particles,''
  Rev.\ Math.\ Phys.\  {\bf 14}, 759 (2002)

\bibitem{bisognano}
  J.~J.~Bisognano and E.~H.~Wichmann,
  ``On the Duality Condition for Quantum Fields,''
  J.\ Math.\ Phys.\  {\bf 17}, 303 (1976)

\bibitem{Moretti1}
C.~Dappiaggi, V.~Moretti and N.~Pinamonti,
  ``Rigorous construction and Hadamard property of the Unruh state in Schwarzschild spacetime,''
  Adv.\ Theor.\ Math.\ Phys.\  {\bf 15}, no. 2, 355 (2011)

\bibitem{Dafermos1}
M.~Dafermos and I.~Rodnianski,
  ``The black hole stability problem for linear scalar perturbations,''
  XVIth International Congress on Mathematical Physics, P. Exner (ed.), World Scientific, London, 2009, pp. 421-433
  
\bibitem{Luk}
J.~Luk,
  ``Improved decay for solutions to the linear wave equation on a Schwarzschild black hole,''
  Annales Henri Poincare {\bf 11}, 805 (2010)
  
\bibitem{Schlue}  
V.~Schlue,
  ``Decay of linear waves on higher dimensional Schwarzschild black holes,''
  Anal. PDE 6 (2013) 515-600
 

\bibitem{Thesis}
S. Hollands, ``Aspects of quantum field theory in curved spacetime,'' PhD Thesis, U. of York (2000)


\bibitem{Dafermos2}
M.~Dafermos, G.~Holzegel and I.~Rodnianski,
  ``The linear stability of the Schwarzschild solution to gravitational perturbations,''
  arXiv:1601.06467 [gr-qc].
  
\bibitem{Chrzanowski}
P.~L.~Chrzanowski,
  ``Vector Potential and Metric Perturbations of a Rotating Black Hole,''
  Phys.\ Rev.\ D {\bf 11}, 2042 (1975).

  \bibitem{Kegeles}
L.~S.~Kegeles and J.~M.~Cohen,
  ``Constructive Procedure For Perturbations Of Space-times,''
  Phys.\ Rev.\ D {\bf 19}, 1641 (1979).


\bibitem{Friederich}
P.~T.~Chrusciel and T.~T.~Paetz,
  ``Characteristic initial data and smoothness of Scri. I. Framework and results,''
  Annales Henri Poincare {\bf 16}, no. 9, 2131 (2015)


\bibitem{Wald2}
R. M. Wald, {\it General Relativity}, U. of Chicago Press (1984)

  
  \bibitem{8} D. L. Jafferis, A. Lewkowycz, J. Maldacena, and S. Josephine Suh. ``Relative entropy equals bulk relative entropy.'' JHEP, 06:004 (2016)

\bibitem{9} X. Dong, D. Harlow, and A. C. Wall, ``Reconstruction of Bulk Operators within the Entanglement Wedge in Gauge-Gravity Duality.'' Phys. Rev. Lett., 117(2):021601 (2016)

\bibitem{10}
T. Faulkner and A. Lewkowycz. ``Bulk locality from modular flow.'' JHEP, 07:151 (2017)

\bibitem{Raamsdonk}
  N.~Lashkari and M.~Van Raamsdonk,
  ``Canonical Energy is Quantum Fisher Information,''
  JHEP {\bf 1604}, 153 (2016)

\bibitem{hiw2}
D.~Garfinkle, S.~Hollands, A.~Ishibashi, A.~Tolish and R.~M.~Wald,
  ``The Memory Effect for Particle Scattering in Even Spacetime Dimensions,''
  Class.\ Quant.\ Grav.\  {\bf 34}, no. 14, 145015 (2017)

\end{thebibliography}
\end{document}